# Effect of contact induced states on minimum conductivity in graphene

Roksana Golizadeh-Mojarad and Supriyo Datta
School of Electrical and Computer Engineering, Purdue Universtiy, West Lafayette, IN-47906, USA.
(Dated: October 14, 2007)

The objective of this paper is to point out that contact induced states can help explain the structure dependence of the minimum conductivity observed experimentally *even if the samples were purely ballistic*. Contact induced states are similar to the well-known metal induced gap states (MIGS) in metal-semiconductor Schottky junctions, which typically penetrate only a few atomic lengths into the semiconductor, while the depth of penetration decreases with increasing band gap. However, in graphene we find that these states penetrate a much longer distance of the order of the width of the contacts. As a result, ballistic graphene samples with a length less than their width can exhibit a resistance proportional to length that is not 'Ohmic' in origin, but arises from a reduced role of contact-induced states. While actual samples are probably not ballistic and involve scattering processes, our results show that these contact induced effects need to be taken into account in interpreting experiments and minimum conductivity depends strongly on the structure and configuration (two- vs. four-terminal) used.

**INTRODUCTION**

Recent experiments [1-12] show that the conductivity of graphene (a single atomic layer of graphite) tends to a minimum value in the range of ~2-12 $e^2/h$ as the Fermi energy $E_f$ approaches the charge neutral Dirac points ($E = 0$) located at the corners of the Brillouin zone. Around these points the density of states is given by $D(E) = (E/(\hbar v_f)^2)(LW/\pi)$ (L: length, W: width) and the conductivity is expected to approach zero as the Fermi energy $E_f$ approaches zero. Consequently, the experimental observation of a non-zero minimum conductivity has stimulated a lot of theoretical work most of which have focused on the carrier scattering mechanisms in graphene [13-17]. However, Beenakker et. al. [18] have shown that even ballistic graphene samples can exhibit a minimum conductivity. The purpose of this paper is to point out that this minimum conductivity in ballistic samples arises from contact induced states and depends strongly on the structure and configuration (two- vs. four-terminal) used. While actual samples are probably not ballistic and involve scattering processes, our results show that these contact induced effects need to be taken into account in interpreting experiments.

Contact induced states are similar to the well-known metal induced gap states (MIGS) in metal-semiconductor Schottky junctions which typically penetrate only a few atomic lengths into the semiconductor, while the depth of penetration decreases with increasing bandgap. However, in graphene we find that these states penetrate a much longer distance of the order of the contacts' width, which seems reasonable since the *graphene acts like a semiconductor with a small gap* that decreases with increasing W.

In this paper, we will present model calculations showing how these contact induced states can help understand many experimental measurements of minimum conductivity in different multi-probe configurations.

*Model:* The basic theoretical model presented here is based on the general non-equilibrium Green's function (NEGF) approach, which has been described elsewhere in detail [19]. The structure is partitioned into channel and contact regions with the channel properties described by a tight-binding Hamiltonian (*H*) appropriate for graphene with a single π-orbital for each carbon atom: all elements of [*H*] are equal to zero except for nearest neighbors for which $H_{mn} = -t; t = 2.71$ eV.

The effect of contacts on the channel is included through the self-energy matrix [$\Sigma_{L,R}$] whose elements are given by $\Sigma_{L,R} = \tau_{L,R} g_{s(L,R)} \tau_{L,R}^+$, where $\tau$ is the coupling matrix between the contacts and channel and $g_s$ is the surface Green's function for the contact. The surface Green's function at energy *E* is obtained from the Hamiltonian for the isolated contact (*H*$_{contact}$) using the relation

$$g_s = [((E+i\eta)I - H_{contact}]^{-1} \quad (1)$$

which is evaluated using a recursive method making use of the tridiagonal nature of $H_{contact}$ [20]. Typically $\eta$ is assumed to be an infinitesimal quantity which for graphene contacts would give a vanishing DOS at $E = 0$. Instead we use a finite $\eta$ adjusted to yield a desired non-zero DOS at $E = 0$ channels being metallic. Figs. 1(c,e) show that the contact density of states penetrates a distance ~ W/2 into the graphene channel.

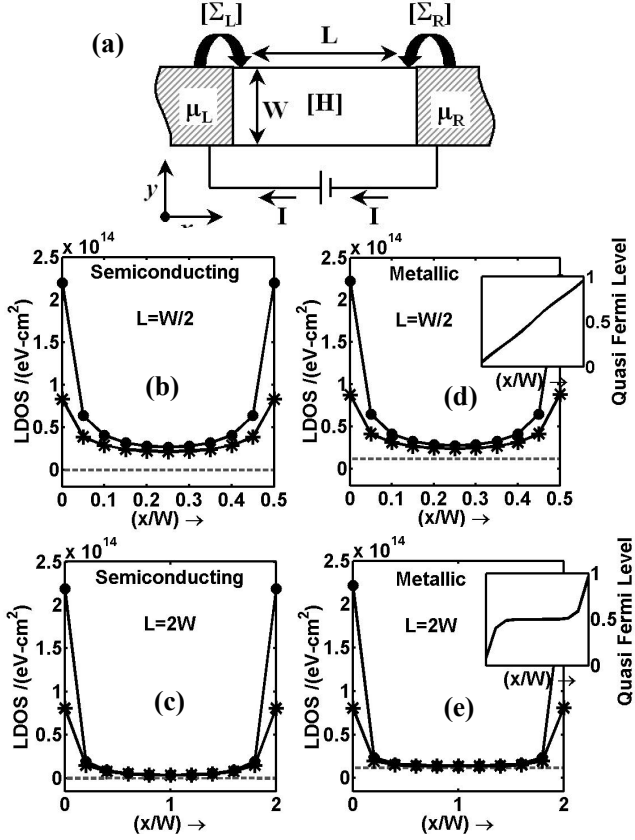

While contact induced states increase the channel DOS, they decrease the channel resistance. For L/W<0.5 resistance increases linearly in a manner reminiscent of Ohm's Law [Fig. 2(a)], *except that this increase is not a result of scattering or momentum relaxation, but from the reduction in "punch-through" of contact induced states.* Nevertheless, if we extract a sheet conductivity using 'Ohm's law' $\sigma = L/RW$, it tends to a constant ($\sim 4e^2/\pi h$) for L/W<0.5 and increases linearly as L/W increased [see ✱-line for low T in Fig. 2(c)]. Our numerical results confirm recent theoretical work by Beenakker et. al. [18] and are in good agreement with the recent experimental measurements [1] as shown later (Tbl. 1). These results are not affected significantly by the precise contact density of states for the range of values studied between [0.8-2.2]×10$^{14}$/(eV-cm$^2$). For large L/W the resistance for metallic channels saturates to $h/2e^2$ corresponding to one spin-degenerate mode [✱-line in Fig. 2(a)], while that for semiconducting channels keeps increasing indefinitely [✱-line in Fig. 2(b)]. Note that this is true at temperatures low enough that one mode is accessible for metallic channels and none for semiconducting channels. More generally, the resistance for long channels saturates to a value $R_S$ that depends on the ratio of the thermal energy $k_BT$ to the in the range [0.8-2.2]×10$^{14}$/(eV-cm$^2$) corresponding to what we estimate for real experimental contacts [1-12].

Fig. 1: Effect of contact induced states on an armchair graphene channel in a general two-terminal configuration shown in (a). General schematic illustration used for NEGF quantum transport calculation. Local density of state (LDOS) averaged over the width plotted against the length (x) for (b) semiconducting channel with L/W=0.5, (c) semiconducting channel with L/W=2, (d) metallic channel with L/W=0.5 and (e) metallic channel with L/W =2. The three curves in each plot represent increasing contact density of states: Dashed line for contact DOS same as channel, ✱-line for contact DOS of 8×10$^{13}$/(eV-cm$^2$) and ●-line for contact DOS of 22×10$^{13}$/(eV-cm$^2$). Also shown in insets for the metallic channel the variation of the quasi Fermi level across the channel normalized to "0" and "1" in the right and left contacts.

***Two-terminal geometry:*** In order to show the effect of contact induced states on minimum conductivity in graphene we start with a simple two-terminal geometry shown in Fig. 1(a). Channels with L/W<1 [Figs. 1(b,d)] show a "punch-through" of DOS from one contact to the other, while in channels with L/W>1 [Figs. 1(c,e)] the DOS in the middle approaches the correct value: zero for semiconducting channels and ~1/$taW$ for metallic channels corresponding to one mode ($a$: carbon-carbon bond length). Similar results are obtained with zigzag graphene, except that there are no semiconducting channels, all energy spacing $kT_W \approx \pi\hbar v_f / 4W$ between successive modes. In general $R_s(T/T_W) = h/q^2 \overline{M}$, where the thermally averaged number of modes $\overline{M} = \int dE M(E)(\partial f/\partial E)$ is approximately equal to $T/T_W + 2$ for metallic channels in our range of interest. Fig. 2 also shows the calculated resistance and the derived conductivity for values of $T/T_W > 1$ indicating the reduction in resistance and enhancement in conductivity expected from the above argument.

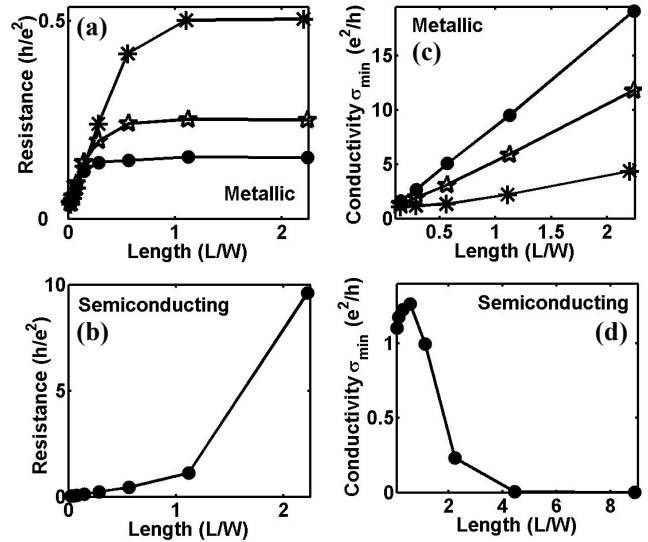

Fig. 2: Effect of the contact induced states on the two-terminal resistance of a graphene channel. Resistance as a function of the channel length for (a) a metallic armchair channel (b) a semiconducting armchair channel. Derived conductivity of the channel as a function of length for (c) a metallic armchair channel (d) a semiconducting armchair channel. ✱-solid line when $T/T_W < 0.5$; ☆-solid line when $T/T_W \approx 3$ and ●-solid line when $T/T_W \approx 6$.

*Spatial variation of the quasi-Fermi level:* Insets in Fig. 1(d-e) show the spatial variation of the quasi-Fermi level for metallic channels. With $L/W = 2$, the quasi-Fermi level is flat in the middle of the channel as one would expect for a ballistic conductor. However, near the contacts it drops linearly. Indeed with $L/W=0.5$ the quasi-Fermi level changes linearly all the way from one contact to the other similar to ordinary diffusive conductors, *although our model includes no scattering*. The linear variation arises simply from the spatial variation in the DOS. As shown in [21], the contact resistance between two regions with "M" modes and "N" nodes is proportional to '$1/N - 1/M$', so that with spatially varying modes described by $M(x) \sim \pi\hbar v_f D(x)$, we expect a potential profile given by $V(x) \sim -d(D^{-1})/dx$, which is approximately linear as shown (inset Fig 1(d)).

*Hall bar geometry:* Many experimental measurements employ a four-terminal Hall bar geometry and one might expect from the flat quasi Fermi level near the middle in inset of Fig. 1(e) that four-terminal measurements would be unaffected by the contact-induced state effects we have been discussing. We now present results that while this is indeed true if the voltage probes are "weakly coupled" to the channel, strongly coupled voltage probes lead to resistance values that are not too different from what we have been discussing for two terminal measurements.

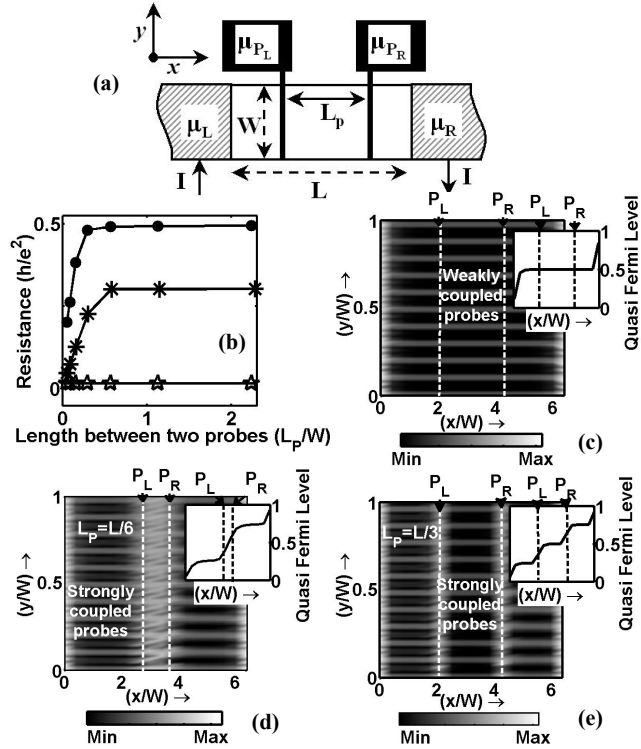

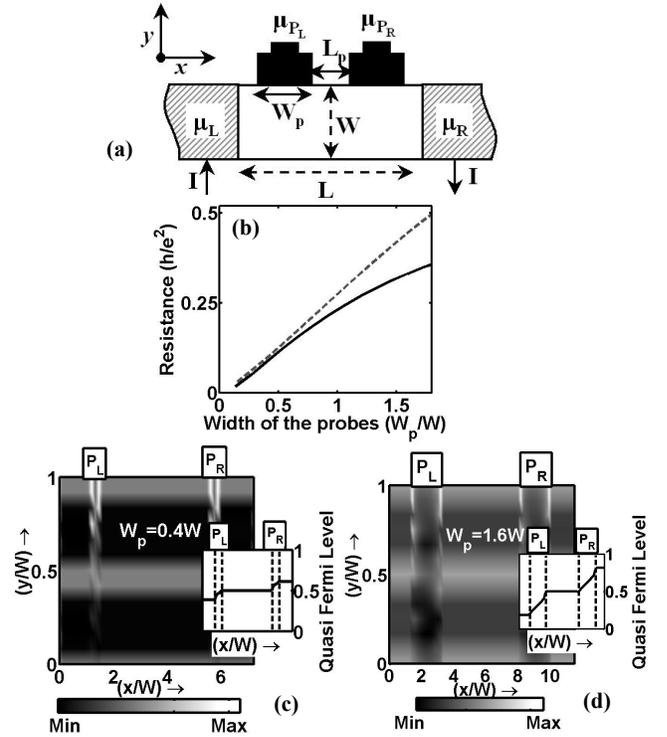

Fig. 3: Effect of contact induced states in Hall bar geometry with full width voltage probes. (a) Schematic illustration of structure. (b) Calculated resistance as a function of the distance between voltage probes ($L_P$). ✩-line for weakly coupled voltage probes; ✱-line for strongly coupled voltage probes with DOS of $2.2\times10^{14}$/(eV-cm$^2$); ●-line for strongly coupled voltage probes with DOS of $0.8\times10^{14}$/(eV-cm$^2$). 2-D Local density of state profile inside a metallic armchair channel for fixed L and W (c) when the voltage probes are weakly coupled, (d) when the voltage probes are strongly coupled for $L_P<W$, (e) when the voltage probes are strongly couple for $L_P>W$. Also shown in insets the quasi Fermi level profile across the channel normalized to "0" and "1" in the right and left contacts.

Fig. 4: Effect of contact induced states in Four-probe structure with side voltage probes. (a) Schematic illustration of structure. (b) Calculated resistance as a function of width of the side voltage probes ($W_P$); solid line for contact DOS of $0.8\times10^{14}$/(eV-cm$^2$), dashed line for contact DOS of $2.2\times10^{14}$/(eV-cm$^2$). 2-D Local density of state profile inside a metallic armchair channel for fixed L, $L_P$ and W (c) $W_P=0.4W$ (d) $W_P=2.4W$. Also shown in insets the quasi Fermi level profile across the channel normalized to "0" and "1" in the right and left contacts.

| Experiment | Device size | Measured $\sigma_{min}$ | Calculated $\sigma_{min}$ |
|---|---|---|---|
| Device 1 W=640nm | $L/W \approx 0.6$ $T/T_W < 0.5$ | $\sim 3\, e^2/h$ | $\sim 2.5\text{-}6\, e^2/h$ |
| Device 2 W=360nm | $L/W \approx 1.06$ $T/T_W < 0.5$ | $\sim 4.5\, e^2/h$ | $\sim 4\text{-}6.5\, e^2/h$ |
| Device 3 W=360nm | $L/W \approx 0.75$ $T/T_W < 0.5$ | $\sim 3.4\, e^2/h$ | $\sim 3\text{-}5\, e^2/h$ |
| Device 4 W=320nm | $L/W \approx 1.44$ $T/T_W < 0.5$ | $\sim 5.6\, e^2/h$ | $\sim 4.8\text{-}7\, e^2/h$ |
| Device 5 W=840nm | $L/W \approx 0.13$ $T/T_W < 0.5$ | $\sim 1.15\, e^2/h$ | $\sim 1\text{-}1.3\, e^2/h$ |
| Device 6 W=400nm | $L/W \approx 0.41$ $T/T_W < 0.5$ | $\sim 1.7\, e^2/h$ | $\sim 1\text{-}2.5\, e^2/h$ |
| Device 7 W=400nm | $L/W \approx 0.58$ $T/T_W < 0.5$ | $\sim 2.25\, e^2/h$ | $\sim 1.9\text{-}3.3\, e^2/h$ |
| Device 8 W=680nm | $L/W \approx 0.14$ $T/T_W < 0.5$ | $\sim 0.98\, e^2/h$ | $\sim 1\text{-}1.6\, e^2/h$ |

Table. 1: Comparison between measured $\sigma_{min}$ and calculated $\sigma_{min}$ for Hall bar geometry of the experiment in Ref. [1]. Since the Hall bar geometry used in the experiment is a combination of side probes and probes with widths less than full, we use dashed line in Fig. 4(b) and ✻-line in Fig. 3(b) to find a range that minimum conductivity can fall in for this type of structure. We assumed $W_P$=250nm as estimated from Fig. 1(b) in Ref. [1]. Note that for the all devices in this table $T/T_W<0.5$.

*Full width voltage probes*: In this geometry [Fig. 3(a)], if voltage probes are weakly coupled to the channel, they do not induce any states in the channel [Fig. 3(c)]. In this case the calculated resistance [✻-line in Fig. 3(b)] is zero, since there is no voltage drop deep inside the channel (ballistic regime) as explained above. Experimental voltage probes, however, are strongly coupled leading to a voltage drop between them as shown in Fig. 3(d,e). The calculated resistance now shows a variation with $L_P/W$ [Fig. 3(b)] similar to the variation of the two-terminal resistance as a function of L/W (Fig. 2). However, with increasing DOS in the probes we find that the resistance goes down [see ✻-line in Fig. 3(b)] which we ascribe to the increased density of contact induced states in the region between the probes.

*Side voltage probes:* With the voltages probes connected to one side of the channel [Fig. 4(a)], the contact induced states from these probes penetrate along the channel width (*y*) rather than length (*x*). If the probe width ($W_P$) is small compared to the channel width, these probes act like weakly coupled full width voltage probes and the resistance is small [Fig. 4(b-c)]. However, with increasing $W_P$ the contact induced states punch through the width of the channel [Fig. 4(d)] and the calculated resistance increases with $W_P$ [Fig. 4(b)]. The distance between side voltage probes ($L_P$) does not affect the measured resistance from this geometry since the contact induced state penetration is along the width of the channel. Higher probe induced density of states cause the probes to affect more of the channel, and hence measure a higher voltage drop/resistance.

| Experiment | Device size | Measured $\sigma_{min}$ | Calculated $\sigma_{min}$ |
|---|---|---|---|
| Device 9 W=1μm | $L/W \approx 0.2$ $T/T_W < 0.5$ | $\sim 1.7\, e^2/h$ | $\sim 1.5\, e^2/h$ |
| Device 10 W=3.3μm | $L/W \approx 0.12$ $T/T_W \approx 1$ | $\sim 2.5\, e^2/h$ | $\sim 1.7\, e^2/h$ |
| Device 11 W=5.74μm | $L/W \approx 0.22$ $T/T_W \approx 1.6$ | $\sim 4.5\, e^2/h$ | $\sim 2\, e^2/h$ |
| Device 12 W=6.15μm | $L/W \approx 0.64$ $T/T_W \approx 1.7$ | $\sim 4.5\, e^2/h$ | $\sim 3.4\, e^2/h$ |
| Device 13 W=7.9μm | $L/W \approx 0.16$ $T/T_W \approx 2.2$ | $\sim 5\, e^2/h$ | $\sim 2.2\, e^2/h$ |
| Device 14 W=1.31μm | $L/W \approx 1$ $T/T_W \approx 0.5$ | $\sim 6.3\, e^2/h$ | $\sim 2.6\, e^2/h$ |

Table. 2: Comparison between measured $\sigma_{min}$ and calculated $\sigma_{min}$ for two terminal geometry of the experiment in Ref. [1]. Minimum conductivity is estimated from Fig. 2(b). Note that for the most of devices in this table, the device width is wide enough to make $T/T_W>0.5$.

***Comparison with experiments:*** Tables 1 and 2 compare experimental measurements from Ref. [1] with our calculated conductivity for *ballistic conductors taking contact induced state effects into account*. Good agreement is seen for four-terminal measurements (Table. 1) but we see discrepancies for several two-terminal measurements (Table. 2) exceed our predicted conductivity. A possible explanation for these discrepancies is the presence of charged impurities (neglected in this discussion) which have been shown to increase the conductivity [6,13-17]. This explanation seems supported by the fact that the discrepancies are limited to wide samples. In summary, we have shown that even ballistic graphene samples with L<W can exhibit a resistance proportional to the length that is not 'Ohmic' in origin, but arises from a decreased rate of contact induced states and this ballistic model can explain many experimental results. Finally, contact-induced states can account for some of the experimental observations such as the structure dependence of minimum conductivity that cannot be explained otherwise.

[1] F. Miao, S. Wijeratne, Y. Zhang, U.C. Coskun, W. Bao, C. N. Lau, Nature **317**, 1530 (2007).
[2] K.S. Novoselov, A. K. Geim, S. V. Morozov, D. Jiang, Y. Zhang, S. V. Dubonos, I. V. Grigorieva, A. A. Firsov, Science **306**, 666 (2004).
[3] K.S. Novoselov, A. K. Geim, S. V. Morozov, D. Jiang, Y. Zhang, S. V. Dubonos, I. V. Grigorieva, A. A. Firsov, Science **438**, 197 (2005).

[4] Y. Zhang, Y. W. Tan, H. L. Stomer, P. Kim, Nature **438**, 201 (2005).
[5] A. K. Geim, K.S. Novoselov, Nature Mat. **6**, 183 (2007).
[6] Y. W. Tan, Y. Zhang, K. Bolotin, Y. Zhao, S. Adam, E. H. Hwang, S. Das Sarma, H. L. Stormer, and P. Kim, arXiv:0707.1807.
[8] J. H. Chen, C. Jang, M. S. Fuhrer, E. D. Williams, M. shigami, arXiv:0708.2408.
[9] S. Cho, Yung-Fu Chen, M. S. Fuhrer arXiv:0706.1597v1.
[10] S. Cho, M. S. Fuhrer, arXiv:0705.3239.
[11] H. B. Heersche, P Jarillo-Herrero, J. B. Oostinga, L. M. K. Vandersypen, A. F. Morpurgo, Nature **446**, 56 (2007).
[12] M. Y. Han, B. Oezyilmaz, Y. Zhang, and P. Kim, Phys. Rev. Lett, 98, 206805 (2007).
[13] K. Nomura, A. H. MacDonald, Phys. Rev. Lett. **98**, 076602 (2007).
[14] Shaffique Adam, E. H. Hwang, V. M. Galitski, S. Das Sarma, arXiv:0705.1540 (2007).
[15] H. Kumazaki, D. S. Hirashima, J. Phys. Soc. Jpn. **75,** 53707 (2006)
[16] T. Ando, J. Phy. Soc. Jpn., **75**, 74716 (2006).
[17] E. H. Hwang, S. Adam, S. Das Sarma. Phys. Rev. lett. **98**, 186806 (2007).
[18] J. Tworzydlo, B. Trauzettel, M. Titov, A. Rycerz, C. W. J. Beenakker, Phys. Rev. Lett. **96**, 246802 (2006)
[19] S. Datta, *Quantum Transport: Atom to Transistor* (Cambridge University Press, Cambridge, 2005).
[20] M. P. L. Sancho, J. M. L. Sancho and J. Rubio J. Phys. F: Met. Phys. **14**, 1205 (1984).
[21] S. Datta, *Electronic Transport in Mesoscopic Systems* (Cambridge University Press, Cambridge, 1997), P. 112